\documentclass[Afour,sagev,times]{sagej}

\usepackage{moreverb}
\usepackage{fancyvrb}
\usepackage[utf8]{inputenc}
\usepackage[dvipsnames]{xcolor}
\usepackage{booktabs} 
\usepackage{graphicx,todonotes}
\usepackage{tabulary}
\usepackage{multirow}
\usepackage{float}
\usepackage{colortbl}
\usepackage{natbib}
\usepackage[perpage]{footmisc}
\usepackage{amsmath}
\usepackage{lscape} 

\usepackage[colorlinks,bookmarksopen,bookmarksnumbered,citecolor=red,urlcolor=red]{hyperref}

\newcommand\BibTeX{{\rmfamily B\kern-.05em \textsc{i\kern-.025em
b}\kern-.08em T\kern-.1667em\lower.7ex\hbox{E}\kern-.125emX}}

\usepackage{listings}
\begin{document} %
\setcounter{secnumdepth}{3}

\runninghead{Candela, G.}

\title{Towards a semantic approach in GLAM Labs: the case of the Data Foundry at the National Library of Scotland}

\author{Gustavo Candela\affilnum{1,2}}

\affiliation{
  \affilnum{1}National Library of Scotland
  \affilnum{2}Centro Biblioteca Virtual Miguel de Cervantes, Universidad de Alicante (Spain)
} 


\begin{abstract} 
GLAM organisations have been exploring the benefits of publishing their digital collections in a wide variety of forms since the 2000s. Many institutions, and in particular libraries, have adopted the Semantic Web and Linked Data principles to their main catalogues. Recent advances in technology and innovative approaches concerning the reuse of the digital collections by means of computational access have paved the way for the creation of Labs within GLAM organisations. In this work, we present a framework to  transform the datasets made available by GLAM organisations under open licenses into LOD. The framework has been applied to three metadata datasets made available by the Data Foundry at the National Library of Scotland. The results of this work are publicly available and can be applied to other domains such as digital humanities and data science. 
\end{abstract}


\maketitle

\section{Introduction}
\label{sec:intro}

During the last decade, technological advances have paved the way for a new context in the GLAM (Galleries, Libraries, Archives and Museums) sector, in which their rich collections are expected to play an important role in the research community. Institutions are adapting to this new environment in terms of services, collections, metadata and skills.\cite{oclc-padilla,oclc-metadata,rluk-manifesto} New approaches are focused on improving the access of the digital collections by using computational methods and increasing datafication based on Data Science, Artificial Intelligence and Machine Learning.\cite{padilla_thomas_2019_3152935, datafication} 

In this context, GLAM Labs have emerged as a new trend addressing the publication under open licenses and reuse of the digital collections in innovative and creative ways.\cite{open_glam_lab} For example, the Data Foundry at the National Library of Scotland publishes data openly and in a variety of file formats,\endnote{\url{https://data.nls.uk/}} LC Labs\endnote{\url{https://labs.loc.gov/}} supports digital transformation at the Library of Congress and KB Lab makes available for the public experimental tools and datasets based on the collections published by the National Library of the Netherlands.\endnote{\url{https://lab.kb.nl/}}

The Semantic Web was introduced early in the 2000s as an extension of the Web based on standards to make it available as machine-readable data.\cite{w3csw} The Semantic Web enables the publication of Linked Open Data (LOD) by using URIs to identify the resources and creating links to external repositories in order to enrich them.\cite{w3cld} Many organisations have explored the benefits of adopting the Semantic Web principles to their catalogues including the Library of Congress,\endnote{\url{https://id.loc.gov/}} the Bibliothèque nationale de France\endnote{\url{https://data.bnf.fr/}} and the Biblioteca Virtual Miguel de Cervantes,\cite{DBLP:journals/semweb/RomeroECS18} amongst others.

While organisations provide digital collections as datasets ready for reuse, these are made available in many cases using traditional standards such as MARC (Machine-Readable Cataloging), less expressive vocabularies such as Dublin Core and text formats such as CSV. In this sense, the application of the Semantic Web principles by GLAM organisations has several benefits including: i) the description of the information with machine-readable, standard and controlled vocabularies; ii) the enrichment with external repositories; and iii) the reuse by third-party actors. In addition, this is a key aspect in order to be part of international and collaborative initiatives such as the Linked Open Data Cloud\endnote{\url{https://lod-cloud.net/}} and Wikidata.

The objective of the present study is to introduce a framework to transform into LOD the datasets made available by GLAM organisations under open licenses following best practices.\cite{padilla_thomas_2019_3152935,matias_frosterus_2021_4572501} The framework was then applied to three metadata datasets made available on the Data Foundry at the National Library of Scotland. The results of this study are publicly available and can be applied to other domains such as digital humanities and data science. 
The main contributions of this paper are as follows: (a) a framework to transform metadata datasets into LOD following best practices; (b) a compilation of Jupyter Notebooks to reproduce the transformation; and (c) the results of the application to a selected set of datasets made available by the National Library of Scotland. These contributions are intended to encourage GLAM institutions to adopt LOD as a key element for publishing their datasets.

The paper is organised as follows: after a brief description of the state of the art in Section~\ref{sec:sota}, Section~\ref{sec:framework} describes the framework employed to transform the datasets into LOD. The application of the method and results are shown in Section~\ref{sec:results}. The paper concludes with an outline of the adopted method, and general guidelines on how to use the results and future work.

\section{Related work}
\label{sec:sota}

The following subsections describe previous approaches of publication of LOD by GLAM organisations, examples of reuse and data quality assessments.

\subsection{Publishing datasets in GLAM institutions}
\label{subsec:publishingLOD}
The publication of digital collections by GLAM institutions has increased during the last decade. These may include different material including maps, images, metadata, OCR text, sound or video. Datasets are available in a wide variety of repositories such as GitHub, Zenodo or dedicated websites. 

In this context, Labs have emerged as a new area in GLAM institutions to make available for the public digital collections under open licenses in order to promote their reuse in innovative ways. Labs provide diverse datasets published under open licenses, prototypes of tools that reuse the contents, APIs, software, collections of reproducible Jupyter Notebooks, etc. In addition, datasets are in general classified by type of material such as metadata, digitised content, maps, etc. For example, the Data Foundry at the National Library of Scotland provides digitised collections, metadata in MARC and Dublin Core, maps and spatial data and organisational information as datasets. Table~\ref{table:labs} shows an overview of GLAM Labs made available by institutions.

\begin{table*}
\caption{Overview of GLAM Labs published by relevant GLAM institutions.}
\label{table:labs}
\scalebox{0.83}{
\begin{tabular}{p{5.9cm}p{11cm}} 
\toprule
Institution & URL \\ 
 \midrule
Austrian National Library & \url{https://labs.onb.ac.at/en/dataset/lod}\\
\hline
Biblioteca Nacional de España & \url{https://bnelab.bne.es} \\
\hline
Biblioteca Virtual Miguel de Cervantes & \url{https://data.cervantesvirtual.com} \\
\hline
Bibliothèque nationale de France & \url{https://api.bnf.fr} \\
\hline
Bibliothèque nationale du Luxembourg & \url{https://data.bnl.lu} \\
\hline
British Library & \url{https://www.bl.uk/projects/british-library-labs} \\
\hline
German National Library & \url{https://www.dnb.de/EN/lds}\\
\hline
Library of Congress & \url{https://labs.loc.gov} \\
\hline
National Library of the Netherlands & \url{https://lab.kb.nl}\\
\hline
National Library of Scotland & \url{https://data.nls.uk}\\
\hline
Royal Danish Library & \url{https://labs.kb.dk}\\
\hline
Royal Library of Belgium & \url{https://www.kbr.be/en/projects/digital-research-lab}\\
    \bottomrule
\end{tabular}
}
\end{table*}

Many organisations have started to explore the benefits of applying the Semantic Web and Linked Data principles to their main catalogues following best practices.\cite{matias_frosterus_2021_4572501} Table \ref{table:lodoverview} shows an overview of LOD repositories made available by relevant GLAM institutions. Previous works have addressed a preliminary analysis of how the transformation could be performed such as the National Bibliography of Scotland.\cite{ifla:nls} While the different controlled vocabularies available and used to describe the bibliographic information share the common goal to help users to find, identify, select, and obtain the resources they need, in some cases, classes were not part of all models or have been renamed.\cite{ala:models} Some examples of controlled vocabularies are:

\begin{itemize}
    \item Bibliographic Framework (BIBFRAME): it was developed by the Library of Congress to produce linked metadata in RDF (Resource Description Framework).\cite{lc:bibframe}
    \item Bibliographic Ontology (BIBO):\cite{bibo} provides main concepts and properties for describing citations and bibliographic references such as books and articles on the Semantic Web.
    \item CIDOC Conceptual Reference Model (CIDOC CRM):\cite{cidoc} it represents an ontology for cultural heritage information. It describes the concepts and relations relevant to the documentation of cultural heritage.
    \item Europeana Data Model (EDM):\cite{europeana} the formal specification of the classes and properties that are used in Europeana. 
    \item Functional Requirements of Bibliographic Records (FRBR):\cite{ifla:frbr} a conceptual model published by IFLA originally in 1997 known as WEMI (Work, Expression, Manifestation, and Item).\cite{code4lib:coyle} The FRBR model was complemented and further developed with FRAD (Functional Requirements for Authority Data) and FRSAD (Functional Requirements for Subject Authority Data). An ontology has been published based on this model.\endnote{\url{https://vocab.org/frbr/core}}
    \item IFLA Library Reference Model (LRM):\cite{ifla:lrm} the original models FRBR, FRAD, and FRSAD were replaced by the IFLA LRM model. It was developed to resolve inconsistencies between the three separate models.
    \item Resource, Description and Access (RDA):\endnote{\url{http://www.rdaregistry.info/}} a compilation of entities and properties as controlled vocabularies based on RDA is available for the public at the RDA Registry as RDA element sets and RDA value vocabularies in RDF.
    \item Schema.org:\endnote{\url{https://schema.org/}} a collaborative approach to create, maintain, and promote schemas for structured data on the Internet.
\end{itemize}

\begin{table*}
\caption{Overview of LOD repositories published by GLAM organisations.}
\label{table:lodoverview}
\scalebox{0.73}{
\begin{tabular}{p{5.9cm}p{3.7cm}p{11cm}} 
\toprule
Institution & Vocabulary & URL \\ 
 \midrule
Austrian National Library & EDM, BIBFRAME, RDA & \url{https://labs.onb.ac.at/en/dataset/lod}\\
\hline
Biblioteca Nacional de España & FRBR & \url{http://datos.bne.es} \\
\hline
Biblioteca Virtual Miguel de Cervantes & RDA & \url{https://data.cervantesvirtual.com} \\
\hline
Bibliothèque nationale de France & FRBR & \url{https://data.bnf.fr} \\
\hline
Bibliothèque nationale du Luxembourg & - & \url{https://data.bnl.lu} \\
\hline
BNB Linked Data Platform & BIBO & \url{https://bnb.data.bl.uk} \\
\hline
Europeana & EDM & \url{https://pro.europeana.eu/page/sparql} \\
\hline
German National Library & BIBFRAME & \url{https://www.dnb.de/EN/lds}\\
\hline
Library of Congress & BIBFRAME & \url{https://id.loc.gov} \\
\hline
National Digital Data Archive of Hungary & DC, FOAF, schema.org, DBpedia & \url{http://lod.sztaki.hu}\\
\hline
National Library of Finland & schema.org, BIBFRAME & \url{https://data.nationallibrary.fi} \\
\hline
National Library of the Netherlands & schema.org, LRM & \url{https://data.bibliotheken.nl}\\
\hline
National Library of Sweden & 
KB Base Vocabulary & \url{https://libris.kb.se/sparql}\\
\hline
Rijksmuseum & EDM, SKOS & \url{https://data.rijksmuseum.nl/controlled-vocabularies}\\
  \bottomrule
\end{tabular}
}
\end{table*}

Other approaches are based on the aggregation of bibliographic catalogues and authority files from several institutions such as the Library of Congress and the national libraries of Norway and Finland.\endnote{\url{https://share-vde.org/}} In addition, innovative linked data editors within the library domain have been developed in order to facilitate the adoption of controlled vocabularies such as BIBFRAME Editor\endnote{\url{https://github.com/lcnetdev/bfe/}} and Sinopia.\endnote{\url{https://sinopia.io/}}

\subsection{Data quality assessment and reuse}
\label{subsec:reuseLOD}
Several aspects can be considered when reusing digital collections such as the license used and the data quality. Previous works are focused on the definition of data quality criteria classified by dimensions to assess LOD repositories.\cite{DBLP:journals/semweb/ZaveriRMPLA16,DBLP:journals/jis/RomeroECS22,DBLP:journals/semweb/FarberBMR18} Many of these criteria are assessed by using SPARQL queries. More advanced approaches are based on the definition of node constraints in the form of Shape Expressions (ShEx) to assess RDF datasets that can be used as additional documentation of the repositories.\cite{DBLP:journals/kbs/Fernandez-Alvarez22} Other initiatives are based on the data quality evaluation of traditional metadata formats for bibliographic records, such as MARC.\cite{DBLP:conf/datech/Kiraly17} Table \ref{table:dataquality} shows an overview of previous approaches to assess the data quality of LOD repositories.

\begin{table}
\caption{Overview of previous works regarding LOD data quality assessment classified by approach.}
\label{table:dataquality}
\scalebox{0.95}{
\begin{tabular}{p{2cm}p{5cm}} 
\toprule
Domain & Approach description \\ 
 \midrule
 Libraries & SPARQL and others\cite{DBLP:journals/jis/RomeroECS22,DBLP:conf/kgswc/Hidalgo-Delgado21}\\
 \hline
  Libraries & ShEx\cite{swebj:candelashex}\\ 
\hline
 General & ShEx\cite{DBLP:journals/kbs/Fernandez-Alvarez22}\\
\hline
 General & SPARQL and others\cite{DBLP:journals/semweb/ZaveriRMPLA16,DBLP:conf/icwe/LangerSGG18,DBLP:journals/semweb/FarberBMR18}\\
\hline
General & Ontology approach\cite{DBLP:conf/ruleml/NayakBL21}\\
    \bottomrule
\end{tabular}
}
\end{table}

Recently, there has been a significant focus on the application of Machine Learning (ML), Computer Vision and Artificial Intelligence to the digital collections published by GLAM institutions.\cite{https://doi.org/10.48550/arxiv.2207.02960, oclc-padilla, quratorSBB, locAI}
Several reuse approaches based on the LOD provided by GLAM institutions addressed the creation of visualisations charts such as maps. Others initiatives such as the RDA Entity Finder\endnote{\url{https://lab.kb.nl/tool/rda-entity-finder}} enable users to browse through the bibliographic Work, Expression, Manifestation and Item entities by using web interfaces. Other approaches have proposed the use of the LOD for Digital Humanities techniques based on knowledge discovery.\cite{DBLP:journals/semweb/Hyvonen20}

With regard to the data enrichment, Wikidata has played a relevant role in GLAM institutions. Properties have been created to link the resources such as authors and works.\cite{DBLP:journals/jis/RomeroECS22} Wikidata follows a collaborative edition approach in which the community is able to enrich the resources. Moreover, Wikidata enables the storage of events, datasets or projects by defining a model to store the information (e.g., the Coding da Vinci cultural hackathons\endnote{\url{https://www.wikidata.org/wiki/Wikidata:WikiProject_Coding_da_Vinci}} and the information about the members and computational access projects of the International GLAM Labs Community). In addition, Wikibase, the software that serves as storage system for Wikidata, has been explored and analysed as a potential tool for cataloguing purposes in libraries.\cite{oclc-wikibase} Other repositories with which to enrich a GLAM dataset may include Virtual International Authority File (VIAF), the GeoNames gazetteer, Getty Vocabularies\endnote{\url{https://www.getty.edu/research/tools/vocabularies/}} or Library of Congress Subject Headings (LCSH).\endnote{\url{https://www.loc.gov/aba/publications/FreeLCSH/freelcsh.html}}

These efforts provide extensive demonstration of how digital collections can be made available for the public and transformed to LOD using different vocabularies and assessment methods. Nevertheless, while previous works are based on the publication of the main catalogues of the institutions as LOD, to our best knowledge, none of the work to date have considered the use of the datasets made available by GLAM Labs. In addition, none of the previous approaches have been focused on the datasets provided by the National Library of Scotland. This analysis will be useful for the GLAM community to identify and agree best practices regarding the publication of datasets as LOD.

\section{A Linked Open Data framework to enhance digital collections in GLAM institutions}
\label{sec:framework}

This section introduces the Linked Open Data framework to enhance digital collections in GLAM institutions. The scope of this study is limited to metadata datasets provided by GLAM institutions under open licenses, such as the ones published in GLAM Labs. The steps of the framework are described below.

\subsection{Selecting a dataset}
The selection of a dataset is the first step and refers to the identification of a digital collection that will be used. GLAM institutions provide digital collections in several locations such as websites, code repository platforms, datasets repositories, etc. Recently, GLAM Labs have emerged as innovative section in GLAM institutions in which open datasets can be found. Relevant aspects to consider when choosing a dataset are for example: i) using known open licenses such as Creative Commons; ii) providing additional documentation such as formats available, provenance information and examples of use; and iii) using a permanent identifier to ease the citation.\cite{doi:10.1177/01655515211060530,padilla_thomas_2019_3152935} 

In addition, data quality is becoming a key element when using advanced methods based on NLP, ML and AI.\cite{DBLP:conf/qurator/Neudecker22,DBLP:conf/icaart/StrienBAHMC20} Recently, carbon footprint information has become a crucial element for example in terms of processing, training AI models, and generation and curation of the datasets that can be additionally considered when selecting a dataset for reuse.\cite{https://doi.org/10.48550/arxiv.2207.02960, Mariette_2022}

\subsection{Data source analysis}
This step includes the preliminary analysis of the original contents provided by the dataset. In some cases, this information might be provided as additional documentation such as the case of the Theatre Posters published by the Data Foundry at the National Library of Scotland.\endnote{\url{https://data.nls.uk/data/metadata-collections/theatre-posters/}} Other types of useful documentation may include research guides about the datasets, such as the case of Chronicling America,\endnote{\url{https://guides.loc.gov/chronicling-america-topics}} and Jupyter Notebooks that explores the content and provides a preliminary idea of the information (e.g., dates, titles, etc.).\cite{nls-doc}

When this information is not available, the difficulty depends on the original data format. For example, CSV files can be easily explored using the Python software Pandas.\footnote{\url{https://pandas.pydata.org}} Other formats based on XML such as MARCXML and Dublin Core may require other tools to extract the contents. Several aspects can be considered for the analysis including:

\begin{itemize}
    \item the total number of resources.
    \item the types of resources included (e.g., writer, maps, works, paintings, etc.).
    \item the number of different metadata fields.
    \item the different values contained in a particular metadata field (e.g., roles or places).
    \item main subjects or topics provided by the contents.
\end{itemize}

When using large datasets in terms of size this step may require an additional task to split the original source or to transform it to an manageable format. For instance, Pandas enables users to explore large data sets efficiently using the \texttt{chunksize} attribute. Other approaches are based on the use of Hadoop and Spark.\cite{10.1145/2814864.2814880}

This process can provide as an output useful information about the size of the dataset and the expressiveness of the vocabularies used.

\subsection{Extracting information}
The extraction of information consists on the identification of the key metadata elements that are relevant to be included in the final dataset. Some examples include title, author, publisher, copyright, sources, etc. Depending on the original format the metadata fields may be accessed in different ways. For example, a title in Dublin Core is provided in a \texttt{dc:title} metadata field while in MARCXML the title is provided by the field 245. 

Several software libraries can be used to extract the metadata such as Pandas, XSLT\endnote{See, for example, \url{https://github.com/vioil/Transforming-MARCXML-with-XSLT/blob/master/marc-to-dc.xsl}} and pymarc\endnote{\url{https://pypi.org/project/pymarc/}} for MARCXML.

Note that when using a transformation tool that automatically generates the RDF output, this step is not strictly required as is described in the next step.

\subsection{Data modelling and transformation to RDF}
The adoption of the Semantic Web concepts requires the use of standard and controlled vocabularies that describe how the information is modelled in terms of classes and properties. RDF provides the basis to model and define the data by using triples.

While best practices recommend reusing existing vocabularies to foster the connection of resources,\cite{ld-bp} there exist tools to facilitate the creation of new vocabularies. \cite{DBLP:journals/aimatters/Musen15} Some examples of vocabularies are introduced in Section \ref{sec:sota}.

When using controlled vocabularies, classes represent a particular type in a domain. For example, the class author represents a writer in a library. Instances of a particular class are identified by URIs that may follow patterns (e.g., \texttt{/author/\{id\}}). Other approaches are based on the use of generic URIs (e.g., identifiers).

The transformation to LOD relies on a mapping process using as input the original sources and generating as an output the metadata described using controlled vocabularies. Each resource is identified by an URI following the patterns established. Then, the different properties are assigned to the resource. 

The transformation can be performed using different methods and tools. For example, many software libraries in several programming languages (e.g., Java, Python or PHP) are available to work with RDF. However, the use of these software libraries requires highly specialised and extensive technical skills. Table \ref{table:transformationlod} presents an overview of methods and tools for working with and creating RDF.

\begin{table}
\caption{Overview of approaches and tools for working with and creating RDF.}
\label{table:transformationlod}
\scalebox{0.90}{
\begin{tabular}{p{2.5cm}p{5cm}} 
\toprule
Tool & Description \\ 
 \midrule
 Apache Jena\endnote{\url{https://jena.apache.org/}} & Java software library\\
 \hline
 BIBFRAME Editor & HTML interface for editing BIBFRAME metadata\\
 \hline
 EasyRdf\endnote{\url{https://www.easyrdf.org/}} & PHP software library\\
 \hline
 marc2bibframe2\endnote{\url{https://github.com/lcnetdev/marc2bibframe2}} & XSLT script to transform from MARC to BIBFRAME\\
 \hline
 OpenRefine\endnote{\url{https://openrefine.org/}} & Transformation interface\\
 \hline
 RDF JavaScript Libraries\cite{rdfJavascript} & RDF JavaScript Libraries\\
 \hline
 RDFLib\endnote{\url{https://rdflib.readthedocs.io}} & Python software library\\
    \bottomrule
\end{tabular}
}
\end{table}

Other approaches are based on the use of tools that transform the original metadata into a specific vocabulary. For instance the Library of Congress provides a set of tools for the transformation from MARC to BIBFRAME.\cite{lc-marc2bib} The use of these tools requires the adjustment of several configuration parameters including the domain name (by default http://example.org), the record that contain the identifier or the serialization format to obtain the output result. Table \ref{table:transformationRDF} shows an overview of previous works addressing the transformation of MARC into RDF controlled vocabularies.

\begin{table}
\caption{Overview of previous works concerning the automatic transformation of the original sources into RDF vocabularies.}
\label{table:transformationRDF}
\scalebox{0.90}{
\begin{tabular}{p{2.3cm}p{6cm}} 
\toprule
Reference & Approach description \\ 
 \midrule
 Marimba\cite{DBLP:journals/lht/Vila-SueroG13} & Convert MARC records to FRBR\\
 \hline
 marc2bibframe2\cite{lc-marc2bib} & Convert MARC records to BIBFRAME\\
 \hline
 MARC2RDA\cite{marc2rda} & Mapping between MARC and RDA\\ 
 \hline
 MINT\cite{DBLP:conf/ercimdl/CharlesITH13} & Cross-domain metadata to EDM\\
 \hline
 marc2dc\footnote{\url{https://github.com/vioil/Transforming-MARCXML-with-XSLT}} & Convert MARC records to DC\\
 \bottomrule
\end{tabular}
}
\end{table}

In addition, there are tools to enable users to edit the data using a specific vocabulary. For example, the BIBFRAME Editor is an HTML user-friendly interface that enables users to edit the metadata.\endnote{\url{https://github.com/lcnetdev/bfe/}}

Recently, mappings between traditional metadata formats and advanced controlled vocabularies have been published.\endnote{See, for example, \url{https://www.loc.gov/bibframe/mtbf/}} In addition, dedicated websites have been published to describe the classes and properties that can be used to describe the information as RDF.\endnote{See, for example,  \url{https://www.iflastandards.info/lrm/lrmer}}
Previous works have addressed the comparison of bibliographic models such as LRM and BIBFRAME.\cite{DBLP:journals/ires/AalbergTM19}

\subsection{Data enrichment}
Linked Data enables the interconnection of resources by using the property \texttt{owl:sameAs}. The use of this property eases the enrichment process by adding contextual information from external repositories. In other cases, new properties are specifically created to add links from a particular dataset. For example, Wikidata provides different properties for each repository.\endnote{See, for example, the property P268 for the Bibliothèque nationale de France.} Section \ref{sec:sota} provides a list of potential repositories to use for enriching a dataset.

In order to identify resources in text and link them to external repositories several techniques can be used. For example, NER (Named Entity Recognition) enable the identification of entities in the text such as authors, organisations and locations. More advanced approaches are based on the identification and linking to repositories such as Wikidata and DBpedia.\cite{DBLP:conf/clef/LabuschN22} An example of entity linking tool is DBpedia spotlight that can be used to automatically annotating mentions of DBpedia resources in text.\cite{DBLP:conf/i-semantics/MendesJGB11} Other software libraries such as spaCy\endnote{\url{https://spacy.io/}} and CoreNLP\endnote{\url{https://stanfordnlp.github.io/CoreNLP/}} enable the creation of pipelines and the training of models for NER and entity linking methods.

\subsection{Quality assessment}
Data quality has become a crucial aspect when publishing a digital collection. Several techniques can be used in this sense based on manual and automatic tasks, and data quality criteria as is described in Section \ref{sec:sota}. 

While SPARQL is a powerful language to query an RDF dataset (e.g., see, for example, Listing \ref{lst:sparql}, more sophisticated approaches are based on the mining of ShEx schemas to define node constraints to be tested against RDF datasets. Listing \ref{lst:shex} shows an example of ShEx schema to validate resources typed as \texttt{foaf:Person} and \texttt{schema:Person}.

\begin{lstlisting} [caption=SPARQL query to retrieve the number of resources containing a subject with the text "Gaelic".,captionpos=b, label={lst:sparql},keywords={},keywords={[2]{}}]
SELECT (COUNT(distinct ?s) as ?total) 
WHERE { 
    ?s dc:subject ?subject . 
    FILTER regex(?subject, "Gaelic")
} 
\end{lstlisting}    

\begin{lstlisting} [caption=Example of ShEx schema to validate resources typed as \texttt{foaf:Person} and \texttt{schema:Person}.,captionpos=b, label={lst:shex},keywords={},keywords={[2]{}}]
ex:Person {
   rdf:type  [foaf:Person]  ;
   rdf:type  [schema:Person]  ;
   foaf:name  xsd:string  ;
   skos:prefLabel  xsd:string  ;
   schema:name  xsd:string
}
\end{lstlisting}

Additional aspects to consider when assessing a LOD repository include the inclusion of labels in multiple languages, the use of a public SPARQL endpoint, the provision of  machine-readable licensing information or the validation of the external URIs.\cite{DBLP:journals/semweb/FarberBMR18, DBLP:journals/jis/RomeroECS22}

\subsection{Publication and Exploration}
The final step corresponds to the publication of the datasets in a public repository. Recently, several platforms have become popular in the research community such as DataCite, Zenodo and Hugging Face. The use of permanent identifiers such as DOI (Digital Object Identifier) is crucial to support the citation.

LOD repositories are in many cases made available using a public SPARQL endpoint. This requires the set up of an environment to install an RDF storage system. Some examples include Virtuoso,\endnote{\url{https://virtuoso.openlinksw.com/}} Jena TDB\endnote{\url{https://jena.apache.org/documentation/tdb/}} and RDF4J.\endnote{\url{https://rdf4j.org/}} 

In addition, RDF datasets can be described by means of controlled vocabularies such as the Vocabulary of Interlinked Datasets (VoID).\cite{w3c-void} This information can be useful for data cataloging and archiving of the datasets as well as to provide additional metadata to users. Listing \ref{lst:void} shows an example of VoID description describing an overview of the properties available. 

\begin{figure*}[t]
\begin{lstlisting} [caption=Example of VoID description for the Moving Image Archive dataset.,captionpos=b, label={lst:void},basicstyle=\small,keywords={},keywords={[2]{}}]
:MovingImageArchive a void:Dataset;
    dcterms:title "Moving Image Archive";
    dcterms:description "RDF data extracted from the Moving Image Archive dataset";
    dcterms:license <https://creativecommons.org/publicdomain/mark/1.0/>;
    dcterms:publisher :NLS;	
    dcterms:contributor :GC;
    dcterms:source <https://data.nls.uk/data/metadata-collections/moving-image-archive/>;
    dcterms:source <https://github.com/hibernator11/nls>;
    dcterms:modified "2022-11-09"^^xsd:date;
    void:feature <http://www.w3.org/ns/formats/Turtle>;
    void:dataDump 
      <https://raw.githubusercontent.com/hibernator11/nls/master/rdf/datasetEnriched.ttl>;
    void:vocabulary <http://xmlns.com/foaf/0.1/>;
    void:vocabulary <http://www.europeana.eu/schemas/edm/>;
    void:vocabulary <https://schema.org/>;
    void:classes 7;
    void:exampleResource <https://example.org/film/0001>;
    void:properties 23;
    void:triples 263476;
\end{lstlisting}
\end{figure*}

Platforms such as the Linked Open Data Cloud enable the inclusion of the LOD datasets in aggregators that can be useful in terms of visibility.

The publication can include examples of use based on prototypes and tools. New approaches have recently emerged to show how to explore a dataset that enable the combination of code and text and can be executed in the cloud based on the use of Jupyter Notebooks.\endnote{See, for example, \url{https://glam-workbench.net/}}

\section{Results}
\label{sec:results}
This section presents the application of the framework proposed in Section \ref{sec:framework} to three datasets provided by the Data Foundry at the National Library of Scotland:

\begin{itemize}
    \item Moving Image Archive (MIA): this dataset represents the descriptive metadata from the Moving Image Archive catalogue which is the Scotland's national collection of moving image.\cite{nls-mia} The dataset is provided as a zip file containing a MARCXML and a Dublin Core files. The dataset is available under a license Creative Commons Zero 1.0 Universal.
    \item National Bibliography of Scotland (NBS): this dataset corresponds to the bibliographic records for the National Bibliography of Scotland and part of an ongoing programme of work to create and expand the bibliography.\cite{nls-nbs} The dataset references materials published in Scotland from National Library of Scotland’s main catalogue.
    \item Bibliography of Scottish Literature in Translation (BOSLIT): it was a Voyager database that was maintained by the BOSLIT Committee including information about Scottish literature in translation which aimed to serve the needs of academic researchers, writers and translators, libraries, schools, literature administrators and general readers.\cite{nls-boslit}
\end{itemize}

We have selected the Data Foundry since it is one of the most relevant GLAM Labs/services providing digital collections as datasets amenable for computational use.\cite{ames_cerl,Ames_2021, ames_sarah_2020_3862050,doi:10.1177/2053951720970576} The datasets have been selected according to the following criteria: i) they are included in the metadata datasets section; ii) they are available under open licenses; and iii) they are available as MARCXML and Dublin Core format.

The vocabularies presented in Section~\ref{sec:sota} were analysed in order to identify the most appropriate vocabulary to describe the metadata provided by each of the datasets. As a result, we used as main vocabularies Schema.org for the Moving Image Archive and BIBFRAME for the National Bibliography of Scotland and BOSLIT datasets. Table \ref{table:overviewlod} presents an overview of the features of the final datasets.

The code repository including a collection of reproducible Jupyter Notebooks is available on GitHub.\endnote{\url{https://github.com/hibernator11/nls-fellowship-2022-23}} Each of the Jupyter Notebooks included in the collection describes a step in the framework applied to the dataset. The LOD datasets generated are available as dump files.

\subsection{Data modelling and transformation to LOD}

The three datasets were transformed to LOD using different techniques and vocabularies as is described below. 

For the Moving Image Archive dataset, the transformation was based on the vocabulary Schema.org, using as main entity the class \texttt{VideoObject}. Schema.org was used since it was adopted by the web community as the shared vocabulary among search engines and provides a rich set of properties to describe multimedia resources such as videos.\cite{DBLP:journals/jodl/FreireRHMI20} Additional classes and properties from EDM and FOAF have been used to describe the resources. Note that the original identifier is used to create the URIs in the RDF dataset (e.g., http://example.org/(filmRef)0002\#Instance)

\begin{table}
\caption{Overview of the RDF datasets generated from the Data Foundry.}
\label{table:overviewlod}
\scalebox{0.95}{
\begin{tabular}{p{3cm}rrr} 
\toprule
Description & MIA & NBS & BOSLIT \\ 
 \midrule
 Classes & 7 & 190 & 124\\
 \hline
 Properties & 23 & 185 & 129\\
 \hline
 External links & 75 & 4274 & 611\\
 \hline
 Triples & 263476 & 47882223 & 4503316\\
    \bottomrule
\end{tabular}
}
\end{table}

\begin{figure*}
\includegraphics[width=15cm]{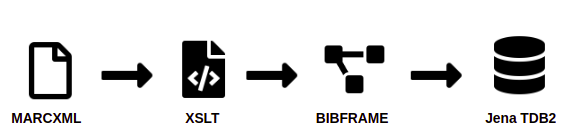}
\caption{Overview of the steps to transform into RDF the National Bibliography of Scotland and BOSLIT datasets using as main vocabulary BIBFRAME.}
\label{fig:nbs}
\end{figure*}

Following previous approaches,\cite{DBLP:journals/ires/AalbergTM19,DBLP:journals/lht/Vila-SueroG13,DBLP:journals/semweb/RomeroECS18} the MARCXML datasets National Bibliography of Scotland and BOSLIT were transformed into BIBFRAME using a XSLT template that was applied to each original record. The final RDF dataset was loaded into a Apache Jena TBS RDF storage system. Figure \ref{fig:nbs} shows an overview of the steps followed.

\begin{figure*}
\includegraphics[width=15cm]{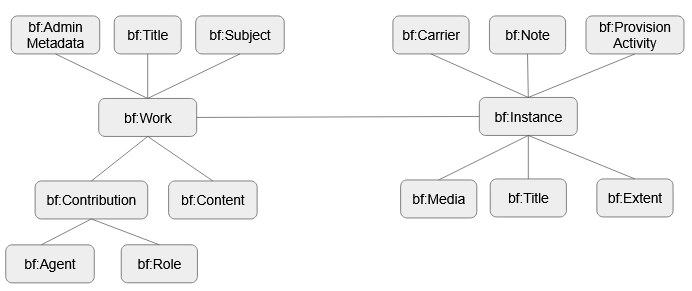}
\caption{Overview of the main BIBFRAME classes used to model data based on the National Bibliography of Scotland.}
\label{fig:nbs-model}
\end{figure*}

The XSLT converter application process the fields of each MARC record and build the two main elements, a \texttt{bf:Work} and a \texttt{bf:Instance}. In addition, the process generates a \texttt{bflc:adminMetaData} property of the \texttt{bf:Work} to include provenance information and generates the \texttt{bf:hasItem} properties of the \texttt{bf:Instance}. URL patterns are based on the baseuri parameter (default http://example.org/), the record ID of the MARC record (by default the value of the 001 field), and a hash URI for the new element. The output is serialized as RDF/XML. For elements that are not typed as \texttt{bf:Work} or \texttt{bf:Instance}, the hash URI is constructed from the element class, the field number, and the position of the field in the MARC record (e.g., 
http://example.org/9923749153804341\#Agent100-12). Figure \ref{fig:nbs-model} shows an overview of the main classes used to model the data based on BIBFRAME.

\begin{figure*}
\includegraphics[width=15cm]{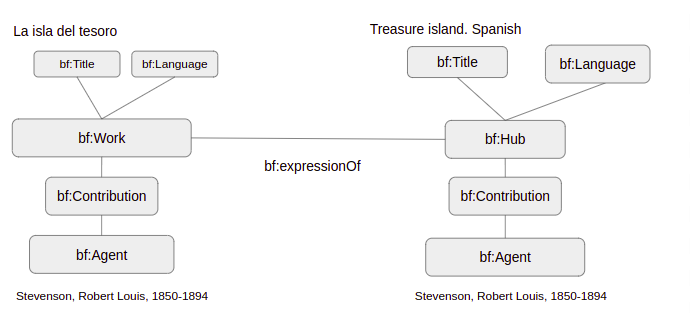}
\caption{Example of BBIFRAME data modelling based on the work \textit{Treasure island}.}
\label{fig:boslit-model}
\end{figure*}

In BIBFRAME, the class \texttt{bf:Hub} is used as an abstract resource that functions as a bridge between two works. MARC metadata fields (e.g., 130 and 240) aiming at storing information about works that has appeared under varying titles may generate an additional resource typed as \texttt{bf:Hub} including the uniform title and the author when available. Figure \ref{fig:boslit-model} shows an example of how the metadata representing translations of the work \textit{Treasure island} in the BOSLIT dataset are modelled according to the BIBFRAME vocabulary.

\subsection{Data quality assessment}
\label{sec:dataquality}

Following previous approaches,\cite{DBLP:journals/semweb/FarberBMR18,DBLP:journals/jis/RomeroECS22,DBLP:journals/kbs/Fernandez-Alvarez22} a validation of the RDF datasets was performed in terms of several dimensions as is described below. Table \ref{tab:farberresults} shows the results obtained.

With regard to the \emph{accuracy} dimension, the \emph{syntactic validity of RDF documents} was assessed based on a random sample of 100 resources per dataset and using the W3C RDF Validation Service.\endnote{\url{https://www.w3.org/RDF/Validator/}} All the documents were assessed as correct.

With regard to the \emph{semantic validity of literals} criterion, regular expressions were used to test textual descriptions included such as roles and text dates. A sample of 100 resources per dataset was tested using a manual revision in order to identify inconsistencies. For example, some inconsistencies in text descriptions concerning roles (e.g., auhtor, orgasnizer, ".", etc.) were identified in the National Bibliography of Scotland dataset. Some textual metadata fields in the Moving Image Archive provide a list of abbreviations (e.g., "a.d." is the abbreviation for "art director") that will require a text processing task to extract all the information. Additional examples of abbreviated roles (e.g., ed.) are included in the BOSLIT dataset.

The criterion \emph{semantic validity of triples} was evaluated using a gold standard as a reference in order to assess whether the metadata was correct. The metadata of a random sample of 50 resources were manually evaluated against VIAF and Wikidata in order to assess whether the metadata information such as title, date of birth, date of death and name was correct. All the datasets satisfied this criterion.

With regard to the \emph{check of duplicate entities} criterion, some potential issues were identified in the National Bibliography of Scotland and BOSLIT RDF datasets:

\begin{itemize}

\item since the RDF transformation is focused on work records, similar authors that potentially are the same are treated as different resources by using different URLs. See, for example, Listing \ref{lst:nbs-stevenson}. The use of external identifiers such as VIAF can help in this sense to facilitate the clustering of records.

\item some BIBFRAME works obtained as output in the transformation process included a \texttt{bf:expressionOf} property linking to a \texttt{bf:Hub} resource. According to the BIBFRAME documentation, a resource typed as \texttt{bf:Hub} represent an abstract resource that functions as a bridge between two Works. For example, MARC fields 130\endnote{\url{https://www.loc.gov/marc/bibliographic/bd130.html}} and 240\endnote{\url{https://www.loc.gov/marc/bibliographic/bd240.html}} store standardised strings such as uniform titles that generates a \texttt{bf:Hub} resource including the main author, the title and the language, when available in the original sources. Since the URL patterns for the \texttt{bf:Hub} and the \texttt{bf:Work} are based on the identifier of the work, the XSLT template may generated repeated \texttt{bf:Hub} resources with the same information but with different URL.

\end{itemize}
 
\begin{figure*}[t]
\begin{lstlisting} [caption=Example of SPARQL query to retrieve the works related with the author Robert Louis Stevenson from the National Bibliography of Scotland.,captionpos=b, label={lst:nbs-stevenson},basicstyle=\small,keywords={},keywords={[2]{}}]
PREFIX bf:<http://id.loc.gov/ontologies/bibframe/>
PREFIX rdfs:<http://www.w3.org/2000/01/rdf-schema#>
SELECT ?label ?a 
WHERE {
  ?s bf:contribution ?c . 
  ?c bf:agent ?a .
  ?a rdfs:label ?label . 
  FILTER regex(str(?label), "Stevenson, Robert Louis")
} 
LIMIT 10
\end{lstlisting}
\end{figure*}

\begin{table*}
\caption{Overview of the results retrieved using the SPARQL query shown in Listing \ref{lst:nbs-stevenson}.}
\label{table:nbs-stevenson}
\scalebox{0.95}{
\begin{tabular}{p{6cm}p{8cm}} 
\toprule
Label & URL \\ 
 \midrule
 Stevenson, Robert Louis, 1850-1894 & http://example.org/9929751083804341\#Agent100-9 \\
 \hline
Stevenson, Robert Louis, 1850-1894 & http://example.org/9923749153804341\#Agent100-12 \\
 \hline
Stevenson, Robert Louis, 1850-1894 & http://example.org/9923749153804341\#Agent800-28 \\
 \hline
Stevenson, Robert Louis, 1850-1894 & http://example.org/9915244463804341\#Agent100-13 \\
 \hline
Stevenson, Robert Louis, 1850-1894 & http://example.org/9944502973804341\#Agent100-10 \\
    \bottomrule
\end{tabular}
}
\end{table*}

Regarding the dimension \emph{trustworthiness}, all the datasets were manually curated in a closed system. While provenance information about the datasets is provided in the Data Foundry, however, no information is available about RDF statements or resources.  

\emph{Consistency} was measured using several SPARQL queries to identify constraints provided by the vocabularies based on statements such as \texttt{owl:disjointWith}. For example, this enables the identification of resources that are typed as Person and Corporate Body. All the datasets satisfied the constraints. In addition, a collection of ShEx schemas has been automatically generated to test against the RDF datasets using the tool sheXer.\cite{DBLP:journals/kbs/Fernandez-Alvarez22} The ShEx schemas can be used as additional documentation for users and curators to better understand how the data is modelled.

Concerning the \emph{relevancy} dimension, none of the datasets supports the ranking of statements, entities or relations, which could be used to, for example, state the order of contributors in a work.

With regard to the \emph{completeness} dimension, the \emph{schema completeness} criterion measures the extent to which classes and relations are included in a dataset. Following previous approaches, a gold standard has been defined with the common classes and properties that could be included based on vocabularies such as BIBFRAME and Schema.org (see Table \ref{table:gold}).

\begin{table}
\caption{Classes and properties selected to assess the completeness criteria.}
\label{table:gold}
\scalebox{0.95}{
\begin{tabular}{p{2cm}p{5cm}} 
\toprule
Pattern & Description \\ 
 \midrule
 Author & name, date of birth, date of death \\
 \hline
 Work/Video & title, date of publication, author \\
 \hline
 Organisation & name\\
 \hline
 Place & name \\
    \bottomrule
\end{tabular}
}
\end{table}

With respect to the \emph{column completeness} criterion, it measures the rate of instances having a specific property, averaged for all the properties shown in Table \ref{table:gold}. The Moving Image Archive dataset obtained a lower score since authors are included in many cases as text descriptions using the metadata field \texttt{schema:creditText}.

With reference to the \emph{population completeness} criterion, it determines the extent to which the datasets covers a basic population. The coverage of resources was compared with a list of resources provided by Wikidata: i) a list of Scottish poets, essayists, novelists and writers for the National Bibliography of Scotland and BOSLIT datasets (see Listing \ref{lst:wikidata-completeness}); and ii) a list of Scottish filmmakers and explorers for the Moving Image Archive.

\begin{figure*}[t]
\begin{lstlisting} [caption={SPARQL query to retrieve Scottish poets, essayists, novelists and writers from Wikidata.},captionpos=b, label={lst:wikidata-completeness},basicstyle=\small,keywords={},keywords={[2]{}}]
SELECT DISTINCT ?sLabel ?viaf
WHERE {
  VALUES ?occupation { wd:Q36180 wd:Q49757 wd:Q6625963 wd:Q11774202}
  ?s wdt:P31 wd:Q5 . 
      ?s wdt:P106 ?occupation .
      ?s wdt:P19 wd:Q22 .
      ?s wdt:P214 ?viaf
      SERVICE wikibase:label { bd:serviceParam wikibase:language "en". }
}
LIMIT 50
\end{lstlisting}
\end{figure*}

\emph{Timeliness} dimension includes several criteria to measure for a digital object the extent to which it is sufficiently up-to-date. The criterion \emph{frequency} measures if the resource includes metadata about when was created, stored, accessed or cited. The frequency of updates was consulted in all the datasets. However, none of the datasets provide information about the validity period or the modification date of statements.

Regarding the \emph{easy of understanding} dimension, user-friendly URLs for the resources are used in all the datasets. For example, Table \ref{table:miaurls} shows the URL patterns used to identify the resources in the Moving Image Archive RDF dataset. All the resources provided a \texttt{rdfs:label} property to describe them and text descriptions were only provided in English. In addition, the datasets used a understandable RDF serialization format such as Turtle. In the particular case of the National Bibliography of Scotland and BOSLIT datasets, the tool marc2bibframe only generated RDF/XML as a result of the transformation using the XSLT template.  

\begin{table}
\caption{URL patterns used to identify the resources in the Moving Image Archive RDF dataset.}
\label{table:miaurls}
\scalebox{0.95}{
\begin{tabular}{p{3cm}p{3cm}} 
\toprule
Pattern  & Description \\ 
 \midrule
 /author/{name} & authors \\
 \hline
 /film/{id} & videos \\
 \hline
 /location/{name} & geographic locations\\
 \hline
 /organisation/{name} & organisations \\
    \bottomrule
\end{tabular}
}
\end{table}

With regard to the \emph{interoperability} dimension, the datasets do not use blank nodes or RDF reification. Only one serialization format is provided for dataset. All the datasets used external vocabularies to describe the metadata including EDM, FOAF, BIBFRAME and Schema.org.

Concerning the \emph{accessibility} dimension, since the URLs are not resolvable due to the use of a testing domain URL, many criteria in this dimension are not satisfied. While the datasets are not available by means of a public SPARQL endpoint, they are available as dump files including metadata.

The \emph{licencing} dimension includes a criterion that measures the extent to which the datasets provide a machine-readable licence. All the datasets satisfied this criterion.

The \emph{interlinking} dimension provides two criteria to measure the external links included in the datasets. The \emph{interlinking via owl:sameAs} criterion computes the rate of instances having at least one external link. In addition to the property \texttt{owl:sameAs}, other properties were identified that link to external repositories such as \texttt{bf:role}, \texttt{bf:language} and \texttt{bf:geographicCoverage}, in particular for the National Bibliography of Scotland and BOSLIT datasets.

With regard to the \emph{validity of external URIs} criterion, the number of HTTP errors were computed for a random sample of 500 URLs shown in triples linking to external repositories. In particular, some issues were identified in the National Bibliography of Scotland dataset regarding the \emph{interlinking} dimension:

\begin{itemize}
\item the automatic transformation process uses the MARC List for Languages controlled vocabulary to create links using the original text information. However, in some cases, the links created are based on non-existent identifiers (e.g., see, for example, the white space in the following URL http://id.loc.gov/vocabulary/geographicAreas/e-uk-\%20st and http://id.loc.gov/vocabulary/languages/d). 

\item resources typed as \texttt{bf:Work} are linked to \texttt{bf:Agent} resources by using a resource typed as \texttt{bf:Contribution} including the name of the author and the role. The Relators terms vocabulary\endnote{\url{https://id.loc.gov/vocabulary/relators.html}} provided by the Library of Congress includes a list of roles that the XSLT template is able to map against the original sources. Up to 15 roles were mapped including, for example, \url{http://id.loc.gov/vocabulary/relators/aut} and 
\url{http://id.loc.gov/vocabulary/relators/ctb}. The XSLT template generated additional roles that were included uniquely as text descriptions instead of URLs (e.g., translator, illustrator, printer, honoree, presenter, etc.).

\end{itemize}

\begin{table*}
\centering
\caption{Summary of results according to the data quality criteria to assess LOD classified by dimensions based on previous works.\cite{DBLP:journals/semweb/FarberBMR18,DBLP:journals/jis/RomeroECS22}}
\label{tab:farberresults}
\scalebox{0.90}{
\begin{tabular}{l p{10cm} l l l l}
\toprule
 Dimension & Criterion & MIA & NBS & BOSLIT \\
 \hline
 \midrule
 Accuracy & Syntactic validity of RDF documents & 1 & 1 & 1 \\
          & Syntactic validity of literals & 0.98 & 0.96 & 0.91 \\
          & Semantic validity of triples & 1 & 1 & 1  \\
          & Check of duplicate entities & 1 & 1* & 1*  \\
 \hline 
 Trustworthiness & On dataset level & 1 & 1 & 1  \\
                 & On statement level & 0 & 0 & 0 \\
                 & Using unknown and empty values & 0 & 0 & 0  \\
 \hline 
 Consistency & Consistency of schema restrictions during insertion of new statements & 0 & 0 & 0  \\
             & Consistency of statements with respect to class constraints & 1 & 1 & 1  \\
             & Consistency of statements with respect to relations constraints & 1 & 1 & 1 \\
 \hline 
 Relevancy & Creating a ranking of statements & 0 & 0 & 0  \\
 \hline
 Completeness & Schema completeness & 1 & 1 & 1 \\
              & Column completeness & 0.61 & 0.79 & 0.69\\
              & Population completeness & 0.35 & 0.5 & 0.3  \\
 \hline 
 Timeliness & Frequency & 0.5 & 0.5 & 0.5 \\
            & Specification of the validity period of statements & 0 & 0 & 0  \\
            & Specification of the modification date of statements & 0 & 0 & 0  \\
 \hline 
 Ease of understanding & Description of resources & 1 & 1 & 1 \\
                       & Labels in multiple languages & 0 & 0 & 0  \\
                       & Understandable RDF serialization & 1 & 1 & 1 \\
                       & Self-describing URIs & 1 & 1 & 1 \\
 \hline 
 Interoperability & Avoiding blank nodes and RDF reification  & 1 & 1 & 1  \\
                  & Provisioning of several serialization formats & 0.5 & 0.5 & 0.5  \\
                  & Using external vocabulary & 1 & 1 & 1  \\
                  & Interoperability of proprietary vocabulary & 1 & 1 & 1 \\
 \hline
 Accessibility & Dereferencing possibility of resources & 0 & 0 & 0 \\
               & Availability of the repository & 0 & 0 & 0 \\
               & Availability of a public SPARQL endpoint & 0 & 0 & 0  \\
               & Provisioning of an RDF export & 1 & 1 & 1 \\
               & Support of content negotiation & 0 & 0 & 0 \\
               & Linking HTML sites to RDF serializations & 0 & 0 & 0 \\
               & Provisioning of metadata & 1 & 1 & 1 \\
 \hline 
 Licensing & Provisioning machine-readable licensing information & 1 & 1 & 1 \\
 \hline
 Interlinking & Interlinking via \texttt{owl:sameAs} & 0.004 & 0,82 & 0,74 \\
              & Validity of external URIs & 1 & 0,66 & 1\\
 \hline
             
  \bottomrule
\end{tabular}
}
\end{table*}

\subsection{Exploration}

\begin{figure} 
\includegraphics[width=\linewidth]{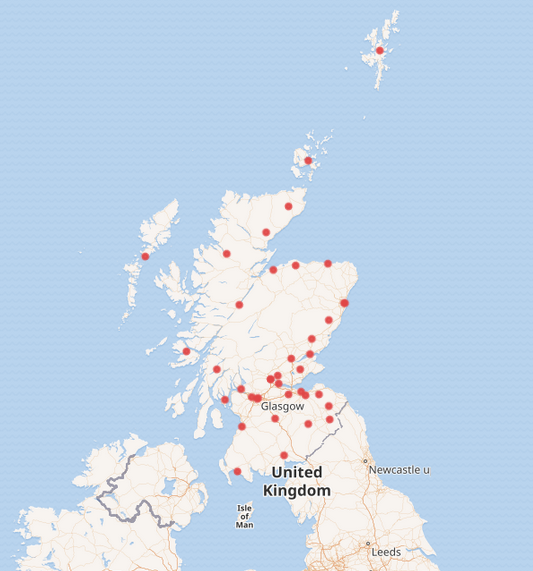}
\caption{Map visualisation created from a SPARQL query in Wikidata using the metadata provided by the Moving Image Archive dataset.}
  \label{fig:map-mia}
\end{figure}

Enriching the data with external repositories such as Wikidata and GeoNames enables the addition of contextual information that can be useful for different purposes including visualisation and data analysis. For example, Figure \ref{fig:map-mia} shows the map visualisation as a result of the SPARQL query that retrieves from Wikidata all the locations provided by the Moving Image Archive dataset.

The transformation process to BIBFRAME included an enrichment step by using the controlled vocabularies provided by the Library of Congress such as MARC List for Languages,\cite{lc_lan} MARC List for Geographic Areas\cite{lc_ga} and MARC Relators terms. In addition, Wikidata provides the property \texttt{P4801} to add identifiers for an item based on the controlled vocabularies maintained by the Library of Congress that can be used to add contextual information.\endnote{\url{https://www.wikidata.org/wiki/Property:P4801}}
Listing \ref{lst:nbs-query} shows and example of SPARQL query to retrieve the works in Spanish from the National Bibliography of Scotland dataset.

\begin{figure*}[t]
\begin{lstlisting} [caption=Example of SPARQL query to retrieve the works in Spanish from the National Bibliography of Scotland.,captionpos=b, label={lst:nbs-query},basicstyle=\small,keywords={},keywords={[2]{}}]
PREFIX bf:<http://id.loc.gov/ontologies/bibframe/>
SELECT distinct ?work ?title 
WHERE {?work bf:language <http://id.loc.gov/vocabulary/languages/spa> .
       ?work bf:title ?resTitle .
       ?resTitle bf:mainTitle ?title 
}
\end{lstlisting}
\end{figure*}

\begin{table*}
\caption{Overview of the results retrieved using the SPARQL query shown in Listing \ref{lst:nbs-query}.}
\label{table:nbs-query}
\scalebox{0.95}{
\begin{tabular}{p{6.5cm}p{8cm}} 
\toprule
URL & Title \\ 
 \midrule
 http://example.org/9944730413804341\#Work & El Palacio de Holyroodhouse\\ 
 \hline
 http://example.org/999356403804341\#Work & La gente y los lugares \\
 \hline
 http://example.org/9929767743804341\#Work & El Ingenioso Hidalgo Don Quixote de la Mancha. (Del Ingenioso Caballero Don Quixote de la Mancha.)\\
 \hline
 http://example.org/9919385013804341\#Work & Una Gramática colonial del Quichua del Ecuador \\
    \bottomrule
\end{tabular}
}
\end{table*}

The BOSLIT dataset is a source of information about Scottish literature in translation. In this way, we explored the dataset in order to analyse how the translations of a work are described using the vocabulary BIBFRAME. Listing \ref{lst:boslit-titles} shows the number of translations in several languages for the works \textit{Treasure island} and \textit{Strange case of Doctor Jekyll and Mister Hyde} and the results are shown in Table \ref{table:boslit-jekyll}. In addition, Listing \ref{lst:boslit-translations} shows a SPARQL query to explore the translation relationships established between the resources in the BOSLIT dataset and the results are shown in Table \ref{table:boslit-editions}.

\begin{figure*}[t]
\begin{lstlisting} [caption={Number of translations in several languages included in the dataset BOSLIT for the works \textit{Treasure island} and \textit{Strange case of Doctor Jekyll and Mister Hyde}.},captionpos=b, label={lst:boslit-titles},basicstyle=\small,keywords={},keywords={[2]{}}]
PREFIX bf:<http://id.loc.gov/ontologies/bibframe/> 
SELECT ?mainTitle (COUNT(?work) as ?total) 
WHERE {
  ?work bf:expressionOf ?exp. 
  ?exp bf:title ?title. 
  ?title bf:mainTitle ?mainTitle
} 
GROUP BY ?mainTitle 
HAVING (?total>1) 
ORDER BY DESC(?total)
LIMIT 20
\end{lstlisting}
\end{figure*}

\begin{table}
\caption{Number of translations per language of the works \textit{Strange case of Doctor Jekyll and Mister Hyde} and \textit{Treasure island} included in the BOSLIT dataset.}
\label{table:boslit-jekyll}
\scalebox{0.90}{
\begin{tabular}{p{6.5cm}r} 
\toprule
Title & No. of works \\ 
 \midrule
Strange case of Doctor Jekyll and Mister Hyde. Italian & 61 \\
Strange case of Doctor Jekyll and Mister Hyde. Spanish & 61 \\
Strange case of Doctor Jekyll and Mister Hyde. French & 56 \\
Treasure island. Spanish & 172 \\
Treasure island. German & 137 \\
Treasure island. Italian & 116 \\
Treasure island. French & 113 \\
Treasure island. Russian & 64 \\
Treasure island. Japanese & 46 \\
Treasure island. Dutch & 42 \\
    \bottomrule
\end{tabular}
}
\end{table}

\begin{figure*}[t]
\begin{lstlisting} [caption={Translations of the work \textit{Strange case of Doctor Jekyll and Mister Hyde. Italian} included in the dataset BOSLIT. Results are shown in Table \ref{table:boslit-editions}.},captionpos=b, label={lst:boslit-translations},basicstyle=\small,keywords={},keywords={[2]{}}]
PREFIX bf:<http://id.loc.gov/ontologies/bibframe/> 
SELECT ?work ?workMainTitle ?exp 
WHERE {
  ?work bf:title ?workTitle . 
  ?workTitle bf:mainTitle ?workMainTitle .
  ?work bf:expressionOf ?exp. 
  ?exp bf:title ?expTitle. 
  ?expTitle bf:mainTitle "Strange case of Doctor Jekyll and Mister Hyde. Italian"
}
LIMIT 20
\end{lstlisting}
\end{figure*}

\begin{table*}
\caption{Editions in Italian of the work Strange case of Doctor Jekyll and Mister Hyde included by the BOSLIT dataset using BIBFRAME  as main vocabulary. The \texttt{bf:Work} is related to a \texttt{bf:Hub} resource describing the uniform title of the main work.}
\label{table:boslit-editions}
\scalebox{0.90}{
\begin{tabular}{p{4.5cm}p{7cm}p{5.5cm}} 
\toprule
Work & Title & Hub  \\ 
 \midrule
http://example.org/15726\#Work & Lo strano caso del dottor Jekyll e del dottor [sic] Hyde ; Il signore di Ballantrae & http://example.org/15726\#Hub240-10 \\
 \hline
http://example.org/9803\#Work & Lo strano caso del dottor Jekyll e del signor Hyde & http://example.org/9803\#Hub240-9 \\
 \hline
http://example.org/9962\#Work & Lo strano caso del dottor Jekyll e del signor Hyde & http://example.org/9962\#Hub240-8 \\
 \hline
 http://example.org/16238\#Work & Il dottor Jekyll & http://example.org/16238\#Hub240-9 \\ 
 \hline
http://example.org/12727\#Work & Il dottor Jekill [sic] & http://example.org/12727\#Hub240-9 \\ 
 \hline
 http://example.org/12333\#Work & Lo strano caso del dottor Jekill [sic] & http://example.org/12333\#Hub240-9 \\
    \bottomrule
\end{tabular}
}
\end{table*}

\subsection{Discussion}
\label{sec:discussion}

Popular and relevant vocabularies often are accompanied of documentation, examples of use, and generation and edition tools. However, while the use of automatic tools eases the transformation process into RDF, there is need for additional customization in terms of features such as URL patterns and the use of external repositories to create links. The process can be constrained by the level of configuration allowed by the tool. In addition, the use of this type of tools relies on the assumption that the original data is of high quality. Moreover, understanding the output of the automatic process, in particular how the data is modelled, may be a cumbersome task. In this way, RDF software libraries can be useful in this context to model the data.

The identification of unique items in an RDF dataset can be improved by including a manual revision. In some cases, the lack of information provided in the original sources (e.g., birth year) can make this process more complex. 

Linked Data promotes the use of URIs to identify resources. Resolvable URIs for the resources require the use of a domain (e.g., https://data.nls.uk/). In this work, we have used the domain https://example.org when evaluating the framework and for testing purposes. In addition, RDF storage systems are required for an efficient access to the data when using large datasets. 

Access and exploration of datasets using traditional metadata formats is constrained due to the use of the textual descriptions. The use of expressive languages to model the data enables the analysis of large datasets using a wide diversity of concepts (e.g., Person, Work, etc.) and properties as access points. In some cases, metadata fields (e.g., roles, dates, etc.) provide textual descriptions that can be further processed to provided a more expressive description of the information. In addition, the use of LOD enables the creation of innovative visualisations that can be useful not only to explore the data but also to gain insight.

\section{Conclusions}
\label{sec:conclusions}

Over the past few years, there has been a growing interest in publishing and reusing the digital collections made available by GLAM institutions. The adoption of the Semantic Web and Linked Data principles provide several benefits to the organisations in terms of interoperability and enrichment.

Based on previous work, we defined a framework described in Section~\ref{sec:framework} for transforming metadata datasets published by relevant GLAM institutions to LOD. The framework was applied to three datasets published by the National Library of Scotland. The evaluation showed that the framework can be useful for other organisations willing to publish datasets as LOD following best practices.

Future work to be explored includes the evaluation of additional datasets, the improvement of the framework to include additional types of datasets such as OCR and the exploration of data spaces to include the final datasets. 

\begin{acks} 
This work has been funded by The National Librarian’s Research Fellowship in Digital Scholarship 2022-23 at the National Library Scotland.
\end{acks}

\theendnotes

\bibliographystyle{SageV}
\bibliography{nls}

\end{document}